# Cell Density Quantification with TurboSPI: $R_2^*$ Mapping with Compensation for Off-Resonance Fat Modulation

Zoe O'Brien-Moran[1,2], Chris V. Bowen[1,2], James A. Rioux[1,2], Kimberly D. Brewer[1,2]

1. Biomedical Translational Imaging Centre, IWK Health Centre & Nova Scotia Health Authority, Halifax, NS.
2. Dalhousie University, Halifax, NS

**Sponsors/Funding Provided by:** NSERC Discovery Grant (KB), NSHRF Establishment Grant (KB), IWK scholarship (ZOM), Brain Canada

**Word Count:** 4131

**Abbreviations Used:** Superparamagnetic iron oxide (SPIO); balanced steady state free precession (bSSFP); single point imaging (SPI); hydroxyethyl piperazineethanesulfonic acid (HEPES); cytotoxic T lymphocytes (CTLs); antigen presenting cells (APCs); compressed sensing (CS).

**Abstract:**

Tracking the migration of superparamagnetic iron oxide (SPIO) labeled immune cells *in vivo* is valuable for understanding the immunogenic response to cancer and therapies. While many sequences are sensitive to SPIO contrast, they lack specificity and provide only semi-quantitative information. Quantitative cell tracking using compressed sensing TurboSPI-based $R_{2*}$ mapping is a promising development to improve accuracy in longitudinal studies on immune recruitment. The phase-encoded TurboSPI sequence provides high fidelity relaxation data in the form of signal time-courses with high temporal resolution. However, early *in vivo* applications of this method revealed that simple mono-exponential $R_2^*$ fitting performs poorly due to the contaminant fat signal in voxels surrounding regions of interest, such as flank tumors and lymph nodes adjacent to adipose tissue. This is especially problematic if there is poor infiltration to the tumor such that immune cells remain near the periphery. The presence of an off-resonance fat isochromat results in modulations in the signal time-course can be erroneously fit as $R_{2*}$ signal decay, thereby overestimating the density of SPIO labeled cells. Simply excluding any voxel with fat-typical modulations results in underestimates in voxels that have mixed content. We propose using a more comprehensive dual-decay ($R_{2f}^*$ and $R_{2w}^*$) Dixon-based signal model that accounts for the potential presence of fat in a voxel to better estimate SPIO induced de-phasing. *In silico* single voxel simulations illustrate how the proposed signal model provides stable $R_{2w}^*$ estimates that are invariant to fat content. The proposed dual-decay model outperforms previous methods when applied to *in vitro* samples of SPIO labeled cells and oil prepared with oil content ≥15%. Preliminary *in vivo* results show that, compared to previous methods, the dual-decay Dixon model improves the balance of $R_{2*}$ specificity versus sensitivity, which in turn will result in more reliable analysis in future cell tracking studies.

**Introduction:**

Immunotherapies are a rapidly growing field of cancer treatments that have had varying degrees of success in the clinic[1,2]. Using imaging to track the migration of immune cells in response to both cancer and cancer therapies is valuable for understanding the underlying immunogenic mechanisms, particularly when used longitudinally. Analyzing immune cell recruitment is especially important when studying immunotherapies, which do not always fit traditional metrics of therapy success[3,4]; alternative metrics based on immune activity may be more relevant.

MRI is one of the most promising modalities for immune cell tracking due to its high resolution, lack of invasiveness and good sensitivity with choice of appropriate contrast agent. Superparamagnetic iron-oxide nanoparticles (SPIO) are the most popular MRI contrast agent for cell tracking[5-8]. SPIO-labeled cells are often imaged using a balanced steady state free precession (bSSFP) sequence[5-9], which is ideal for its high SNR and sensitivity to SPIO effects[10].

Unfortunately, there are well-documented challenges associated with tracking SPIO-labeled cells[11]. For example, specificity issues arise when endogenous sources of negative contrast, such as tumor necrosis (*Figure 1 from* [12], *detailed methods in supplementary methods*), are misinterpreted as the hypo-intense signal voids associated with clusters of labeled cells. Accounting for false positives requires integration with biological techniques (via tumor necropsy and histochemistry) to validate findings for each subject, which is difficult to scale to large studies and eliminates the possibility of longitudinal studies.

Quantification is another challenge for MRI cell tracking applications since sequences like bSSFP enable only qualitative analysis of changes in image contrast. The analysis can become *semi-quantitative* by comparing the signal intensity histograms before and after labeled cell injection[13] or using a matched control[7], but these methods offer only relative quantification. Relaxation rate mapping is a popular method for attempting to quantify SPIO labeled cells, since $R_2^*$ ($R_2^* = 1/T_2^*$) increases linearly as a function of cell density[14,15]. However, many $R_2^*$ mapping techniques proposed to date are unable to resolve rapidly decaying signals restricting quantitative dynamic range, and prone to blooming artifacts induced by the very species which they are intended to quantify[16,17].

Fluorine-19 ($^{19}F$) based methods address some of these issues[18-20], but the gains are countered by a significant decrease in sensitivity compared to proton MRI [19], thereby requiring far more injected cells, larger numbers of cells being recruited to regions of interest, or a higher concentration of $^{19}F$ to be localized within cells.

Previous preclinical work done by our lab demonstrated the use of 3D TurboSPI, adapted for quantifying labeled cells[16,17]. TurboSPI is an accelerated multi-echo single point imaging (SPI) sequence that was first developed for porous media with short $T_2^*$ relaxation times (~1ms). This sequence produces high temporal resolution data (sampled at ~100 kHz) which enables high fidelity $R_2^*$ mapping with greater dynamic range and fewer artifacts due to the purely phase-encoded acquisition[16,17,21,22].

While the initial TurboSPI studies proved promising, a pilot preclinical *in vivo* study that compared multiple immunotherapy treatment groups[12] revealed a new potential source of error: the presence of off-resonance fat modulations in the signal time course. Since the original analysis used a simple mono-exponential model to estimate the $R_2^*$ relaxation parameter, fat signal modulations can falsely indicate elevated signal decay, resulting in apparent regions of elevated R2* decay falsely appearing as elevated SPIO-labeled cell density. That is, one specificity issue was traded for another in the form of false positives from fat. These errors are especially problematic near the fat pads at the tumor periphery and lymph nodes, areas of potentially high cell recruitment (*Figure 2 from* [12], detailed methods in supplementary methods). The larger voxels necessitated by SPI can result in mixed content of fat and SPIO labeled cells, such that the associated time course will reflect both types of signal evolution. Masking out voxels whose time course contains modulations at the typical lipid frequency (447 Hz) will eliminate fat-contaminated voxels from the map, but this leads to underestimates in cell numbers.

More robust analysis to accommodate voxels containing both labeled cells and fat is required. In this work, we propose a modified Dixon-based signal model that simultaneously estimates $R_2^*$ and fat fraction to improve fitting accuracy, such that $R_2^*$ can be quantified independent of the fat content within a voxel. The converse problem has been investigated previously in the literature; for example, Yu[23], O'Regan[24], Chebrolu[25] and Reeder[26] obtained accurate information of fat content in the liver, regardless of $R_2^*$ effects from hepatic iron. However, this is the first attempt to address the problem of improving the accuracy of $R_2^*$ estimates in TurboSPI in the presence of fat.

This work begins with an *in silico* investigation to characterize the problem, an *in vitro* experiment to validate the improved fitting scheme on real data, and finally a preclinical *in vivo* test on a mouse injected with SPIO labeled cells.

**Experimental**:

*Models Used:*

Three fitting models/techniques were compared in these experiments. The simplest method is a mono-exponential single species (water) decay model (Eq. 1), or "single-decay" technique. This method simply fits the magnitude portion of the signal decay as

$$S = We^{-\Delta t/T_2^*}$$ [Eq. 1]

W is the signal amplitude for the water isochromat and, $T_2^*$ is the decay rate of water. $\Delta$t is the difference in time between the centre of the spin-echo peak and the data acquisition. The second technique attempts to address fat voxels by performing a post-acquisition fat exclusion combined with the single-decay model, henceforth referred to as the "exclusion" technique. During the fitting process, the algorithm identifies modulations in the time-course that have fat-specific periodicity, and excludes these voxels from the analysis, i.e. sets $R_2^*$ = 0 for each of those voxels. For the exclusion criteria, fat is identified when the signal phase changes non-uniformly around TE + 1.1ms, i.e. when the fat and water signals are out of phase. This technique was previously applied to *in vivo* data but is unable to address "mixed voxels" containing both fat and SPIO labeled cells[12].

The third technique is the proposed model: modified dual-decay Dixon, or the "dual-decay" technique, which simultaneously estimates the fat fraction as well as separate $R_2^*$ relaxation rates for fat and water (Eq. 2).

$$S = We^{i(\omega\theta_w + \theta_\tau\Delta t)}e^{-\Delta t/T_{2w}^*} + Fe^{i(\theta_f + \Delta\omega_{wf}\Delta t)}e^{-\Delta t/T_{2f}^*} \quad \text{[Eq. 2]}$$

W, F, $\theta_w$ $\theta_f$ are signal amplitudes and phases for the water and fat isochromats, respectively, $T_{2w}^*$ and $T_{2f}^*$ are separate decay rates for the water and fat species, $\Delta\omega_{wf}$ is the frequency offset between the isochromats, and $\theta_\tau$ is a slow time dependent phase change due to the SPIO[26,27].

Dual-$T_2^*$ Dixon models have been used in the literature for other purposes. Chebrolu[25] and O'Regan[24] investigated using independent $T_2^*$ estimates to improve fat quantification, and Horng[28] and Reeder[26] compared methods that use $T_2^*$ to correct fat quantification. Replacing a common $T_2^*$ with independent fat and water $T_2^*$ values has been debated in the literature. Chebrolu found that using dual $T_2^*$ correction methods reduces error, especially in phantoms with high levels of both SPIO and fat[25] since the SPIO has a greater $T_2^*$ effect on water than fat.

*In Silico Investigation and Model Development:*

*In silico* analyses were performed to investigate the combined effects of $R_2^*$ decay and fat modulations. All aspects of the *in silico* work were implemented using customized in-house Matlab 2017b code (Mathworks, Natick, MA).

The TurboSPI signal was first simulated without fat contributions using Monte Carlo methods, which are described in more detail elsewhere[28-30]. Briefly, water molecules diffuse through a randomly distributed grid of spherical magnetic field perturbers of defined size, susceptibility and volume fraction. The magnetic field experienced by each proton is calculated at each time step to track the accumulated phase. The echo time (TE), repetition time (TR), and sampling rate were matched to the TurboSPI sequence used for *in vitro* and *in vivo* experiments. The result is a complex magnetization time course computed from the excitation pulse through the spin echo to t=3TE/2. N=30 000 independent simulations were summed to give the final signal, an example of which is shown in Figure 3 with the typical acquisition window (for *in vitro/in vivo* data) highlighted. $T_2$ relaxation was simulated to be long (1s), a reasonable simplification since $T_2 \gg T_2^*$ for SPIO compartmentalized in cells[29].

In this work, a simplified single peak lipid model with the common chemical shift of 3.5 ppm ($\Delta f_{wf}$ =447 Hz at 3T) is added to the simulations. The signal from fat ($S_f$ = e^{-i2$\pi$($\gamma$Bo-$\Delta f_{wf}$)t} was added to the simulated TurboSPI signal for various fat fractions (ff = 0 to 100% in steps of 5%) to represent lipid contributions (*Figure 3*). Rather than estimating fat fraction separately to correct the $R_2^*$ measurement, both parameters are estimated simultaneously using the dual decay model [Eq. 2].

Presently, the fit is only performed on the portion of the decay from t = 10.5 ms to t = 12.5 ms. This is to avoid the spin echo peak at TE = 10 ms, near which there exists a deviation from the model's predicted exponential decay (the presence of diffusion gives rise to non-exponential signal attenuation[30]). Away from t=TE, this additional factor approaches unity and the proposed model becomes valid.

The real and imaginary channels were concatenated and fit as a single unit to accommodate bounded optimization using Matlab's *lsqcurvefit*, which does not support bounds for complex inputs. Data were simulated and fit with a range of fat fractions from 0 to 1, to demonstrate stability in the $T_{2*}$ measurement, and with a range of perturber volume fractions to ensure stability with respect to cell density. Estimated fat fractions were compared to the known input, and estimated $T_{2*w}$ values were compared to the gold standard value obtained from fitting a mono-exponential decay with no added fat.

*In Vitro Experiment*

Images were acquired using phantoms prepared with equal concentrations of SPIO-labeled cells, but varying fat fractions and two "no cell" controls. Cells were isolated from C57BL/6 mice obtained from Charles River Laboratories (St. Constant, PQ); all procedures adhered to approved ethics protocols for animal care according to the University Committee on Laboratory Animals, Dalhousie University. CTLs were harvested from the inguinal, mesenteric, brachial, axillary, and submandibular lymph nodes from C57BL/6 mice, cultured according to internal protocols over 8 days, and loaded with Rhodamine B SPIO nanoparticles (30 nm, Biopal, Worcester MA) to a concentration of 5 pg/cell using passive *in vitro* incubation for 24 hours.

Solutions of SPIO-labeled cells, water, and peanut oil were suspended in a polyacrylamide gel and placed in 5 mm NMR tubes. The gel contained 10% sodium dodecyl sulfate, 40% acrylamide, 1M hydroxyethyl piperazineethanesulfonic acid (HEPES), 10% ammonium persulfate, and tetramethylethylenediamine. Nine such phantoms were created; seven with SPIO labeled cells (fat content = 0, 5, 10, 15, 20, 30, 40 %) and two controls with no cells (20, 40% fat). The phantoms with SPIO labeled cells contained a constant concentration of $2 \times 10^6$ SPIO loaded CD8+ T lymphocytes (CTLs) per 1 mL solution.

All MR data were acquired using a 3 Tesla horizontal bore pre-clinical MR scanner (Oxford, UK, console by Agilent, Santa Clara, CA). For imaging, NMR tubes were placed inside a cylindrical MR-transparent phantom holder filled with doped water (160 $\times 10^{-6}$ M $MnCl_2$). Samples were imaged in sets of three using a 2D TurboSPI sequence[17] (TR=250 ms, $TE_{effective}$ = 10 ms, ETL=8, ESP=10ms, matrix = 128x128, FOV=50 x 50 $mm^2$, scan time 8.5 min) with and without a chemically-selective fat saturation pulse (90° sinc). Signal time course data were fit for each voxel using the exclusion method[12], and dual-decay model (Eq. 2). When fitting *in vitro* data, the time dependent variables θτ and $\Delta\omega_{wf}$ also incorporate the effects of Bo inhomogeneity, without the need for a separate fitting parameter. While both $T_{2w}*$ and $T_{2f}*$ are estimated, it is $R_{2W}* = 1/T_{2w}*$ that is reported in the final maps.

*In Vivo Experiment*

On Day 0, C57BL/6 mice received a subcutaneous injection of C3 cells from the murine cervical cancer C3 cell line[7,31,32] ($5 \times 10^5$ cells in 100 μL in the left flank).

CTLs were again harvested from the inguinal, mesenteric, brachial, axillary, and submandibular lymph nodes of disease matched donor mice (implanted *Day -7*), isolated and cultured according to internal protocols (*Day 12-Day 19)*. CTLs were primed, using antigen presenting cells (APCs) at a ratio of 1:10 APCs:CTLs, to recognize the R9F antigen of C3 cells[7], thereby promoting cellular recruitment. The CTLs were loaded with SPIO nanoparticles on *Day 19* (30 nm, Biopal, Worcester MA) and $8 \times 10^6$ SPIO labeled CTLs were delivered via tail vein injection on *Day 20.* The animal subjects were imaged 24 hours after injection to allow for biological uptake.

A high-resolution anatomical scan was performed first using bSSFP (TR = 8ms, TE = 4ms, FA = 30°, 4 phase cycles, matrix = 256 x 170 x 170, FOV = 38.4 x 25.5 x 25.5 mm$^3$, scan time 64 min). Signal time course data were provided by a 3D TurboSPI scan accelerated by compressed sensing[16] (TR=250 ms, TE$_{effective}$ = 10 ms, ETL=8, ESP=10ms, matrix = 96 x 96 x 48, FOV=30 x 30 x 30 mm$^3$, acceleration factor = 8, Fat Sat = 90° sinc, scan time 29 min). A matched parameter fast spin echo (FSE) pre-scan (scan time 2.5 min) preceded the 3D TurboSPI sequence to guide the pseudo randomly under-sampled acquisition and constrain the final compressed sensing (CS) reconstruction.

After the CS reconstruction was performed using an in-house Matlab program[16], the time course data were fit for each voxel using three techniques: single-decay (Eq. 1), exclusion (as in *in vitro*) study, and dual-decay (Eq. 2). 3D bSSFP and $R_2$* data were co-registered rigidly using the FSE scan as an intermediary and fused in VivoQuant (inviCRO, Boston MA) for visualization.

**Results**:

*In Silico*

Figure 4 gives an example of the fitted time course for one set of parameters. Though the split real-imaginary function (Figure 4b) is less visually intuitive as a signal time course than the complex equivalent (Figure 4a), this allows a bounded fit that provides better performance for noisy data while still matching well with the simulation (RMSE = 9.4 x 10$^{-4}$),

The accuracy of fat estimation over a range of simulated fat fractions is shown in Figure 5a, while Figure 5b shows the stability of the water T$_2$* parameter estimate across the range of fat fractions for three different perturber volume fractions. When fit properly, the water and fat terms are separated appropriately, and T$_{2w}$* is invariant to fat content while changing appropriately at different volume fractions (i.e. different cell densities). This stable behaviour is observed from 0-99% fat (with a slight deviation at fat = 5%) suggesting that the dual-decay model is stable for mixed voxels, but the fit deteriorates at 100% fat since the model is no longer appropriate. Standard deviations (σ) of the T$_{2*}$ measurements from 0-99% fat are σ =0.23ms (ζ = 1.0 x 10$^{-5}$), 0.16ms (ζ = 1.5 x 10$^{-5}$), and 0.12ms (ζ = 2.0 x 10$^{-5}$), reflecting how fit performance improves with increased R$_2$* effect.

White Gaussian noise was added to simulated TurboSPI data to investigate noise performance of the proposed dual-decay model. This model was evaluated for 20 different repetitions each at five fat fractions, with temporal signal-to-noise ratios (tSNR) of 1 to 40. The mean and standard deviation of the resulting fat fraction and T$_2$* estimates are reported in Figure 6. While the dual-decay model exhibits poor predictive performance for very low tSNR, the mean estimates plateau at tSNR ~ 20. An *in vivo* TurboSPI scan typically has tSNR >40.

*In Vitro*

$R_2^*$ maps calculated using the dual-decay model were compared to those calculated using the exclusion method (mono-exponential decay while eliminating voxels with fat periodicity). Ideally, the "no cell (NC)" tubes should have low $R_{2w}^*$, the mixed fat/water tubes should have equal $R_{2w}^*$ to reflect their equal cell densities, and fat saturation should not affect $R_{2w}^*$. Figure 7 shows example $R_{2w}^*$ maps for the phantoms. The exclusion technique without fat saturation removes all voxels with fat content > 15%. While this is valid for the NC tubes, it is inaccurate for those tubes with mixed voxels, since the $R_{2w}^*$ values used to assess cell density are not computed. The use of fat saturation appears to further reduce accuracy due to incomplete removal of fat signal. The dual-decay technique results in more stable $R_{2w}^*$ estimates for the mixed tubes and predicts lower $R_{2w}^*$ values for the NC controls, although ideally these two tubes would have even lower estimates.

The findings for the fat/water/cell mixture phantoms are summarized in Figure 8, where the $R_{2w}^*$ estimates are plotted against the known fat fractions to test the metric's invariance to fat content. Here we show parameters estimated using a common-decay (where $R_{2w}^* = R_{2f}^*$) model alongside the dual-decay and exclusion techniques. The dual-decay gives the most stable $R_{2w}^*$ estimates across most fat fractions, while the reliability of the exclusion method deteriorates for fat ≥15%, and the estimated $R_2^*$ from the common-decay model decreases with increased fat content. We note that acquiring data with a fat saturation pulse results in slight $R_{2w}^*$ underestimates from the dual-decay model, and wildly inconsistent $R_{2w}^*$ estimates from the exclusion technique.

*In Vivo*

Finally, the dual-decay model was tested on *in vivo* data and compared to the single-decay and exclusion techniques. As shown in Figure 9, the proposed dual-decay technique exhibits fewer erroneously fit voxels than the single-decay model, but more than the exclusion method. However, the strict conditions imposed by the exclusion technique results in drastic underestimates for potentially mixed voxels (as illustrated in the *in vitro* data) and, therefore, no $R_{2w}^*$ values above threshold are seen in the tumor or adjacent lymph node, which is not consistent with the known behavior of these loaded cells.

**Discussion**

The proposed comprehensive dual-decay Dixon model successfully provides accurate estimates of $T_{2w}^*$ despite changing fat content in simulated TurboSPI data. Simultaneously estimating these parameters improves accuracy in both fat fraction and $T_2^*$ measurements, though only the $T_2^*$ measurement is the focus of this work. We investigated $T_{2w}^*$ stability with respect to fat content for different levels of physical complexity: data that neglected $T_{2f}^*$, data with a common $T_2^*$, and finally for the most physically descriptive case of independent $T_{2w}^*$ and $T_{2f}^*$. The $T_{2w}^*$ values estimated using the dual-decay model compared favorably to gold standards obtained from using a simple mono-exponential fit on a simulated time course with no added fat. The *in silico* results suggest that the proposed model should give $T_{2w}^*$ estimates that are invariant to voxel fat content and thus result in more accurate and descriptive $R_2^*$ maps.

A limitation of the *in silico* work was that it neglects the effects of unintended (i.e. not from SPIO) $B_0$ inhomogeneity to investigate the fit response. However, in translating the model to non-idealized data, we allowed $\Delta\omega_{wf}\Delta t$ and $\theta_\tau\Delta t$ to vary to reflect the time-dependent $B_0$ phase effect ($\phi=\gamma\Delta B_0\Delta t$). This should be sufficient for the current investigation, since the goal is not to explicitly decouple or map these parameters, but to consider their combined effect.

Reeder and Horng[26,28] acknowledge that a dual-decay model provides a more accurate description of the physical system, but is hindered by poor noise performance from adding more fit parameters. When using a common $T_2^*$, as suggested originally by Glover[27], if $T_{2f}^* \neq T_{2w}^*$, the common $T_2^*$ will be an average that changes with fat content, which is detrimental to quantification of $T_2^*$. We tested this in Fig. 8 but found that, in the *in vitro* and *in silico* experiments, using the more physically descriptive "dual" $T_2^*$ model[25,26] gave more accurate results, even though it added another parameter. Naturally, increasing the number of unknowns necessitates an increase in the number of measurements, but TurboSPI data (simulated or otherwise) are well equipped to provide this since ~200 points are routinely used in the fitting region for real data. A noise performance analysis using simulated Gaussian noise indicates that fit results stabilize at tSNR = 20 for both common and separate $T_2^*$.

When applied to *in vitro* data, as predicted, the dual-decay signal model provided $R_{2w}^*$ estimates that were more resistant to changes due to high fat content in mixed voxels, as compared to the exclusion technique or common-decay model. The benefits of using a more comprehensive model becomes apparent when fat content rises above ≥15%; both techniques were similarly stable in their estimates for fat <15%, but the standard deviation of the exclusion technique increased at 15%, and accuracy deteriorated further for higher fat contents. When using a common decay rate, we note that the $R_2^*$ estimates decrease linearly with increased fat content, reflecting a mixing of the two different relaxation rates.

Fat saturation had a significant impact on $R_{2w}^*$ estimates when using both the exclusion and common-decay techniques. Without fat saturation, mixed voxels with fat content >15% were removed from the map calculated using the exclusion technique, meaning that any SPIO-labeled cells present will be ignored. After fat saturation, the exclusion method gave over-estimates of $R_{2w}^*$ for 20% and 30% fat and eliminated mixed voxels with fat content of 40%. When fat is present in small quantities, the fat saturation pulse suppresses enough of the fat signal to evade exclusion, but the remaining fat signal still influences the overall amplitude near the spin echo, resulting in overestimated $R_{2w}^*$. The $R_{2w}^*$ estimate should ideally be invariant to a voxel's fat content, and these *in vitro* data demonstrate the improved accuracy of the proposed dual-decay technique. In addition to being more invariant to fat content, the $R_{2w}^*$ measurements from dual-decay were also less affected by the fat saturation pulse. However, while the effect was not as severe, fat saturation did consistently result in slight underestimates of $R_{2w}^*$. These underestimates were likely due to SPIO induced line broadening in the sample and partial saturation of the water peak, even when using a spectrally selective fat saturation pulse.

*In vitro* findings suggest that the dual-decay model should result in fewer misfit voxels, improving *in vivo* specificity while maintaining sensitivity to mixed voxels. Other techniques could assist in further discriminating SPIO labeled cells from pure fat; for example, future implementations may consider an adaptive approach that uses fat content to prescribe the fitting technique on a voxel-by-voxel basis. That is, the dual decay model could be used as a first-pass fitting method, with the single decay model re-applied to voxels below a certain fat threshold, and with voxels above a certain fat threshold eliminated. This adaptive approach would provide the benefits of each technique.

The present work uses a simplified lipid spectral model. Comprehensive spectral modeling is less crucial to this work than in applications that require more accurate fat quantification. The reasoning is similar to those fat-focused publications which note that, while using separate $R_{2w}^*$ and $R_{2f}^*$ values is more physically accurate, the benefit is not worth the additional model parameter. Since this work focuses on accurate $R_2^*$ quantification, multi-peak fat spectra may not be beneficial enough to be worth complicating the model. Nonetheless, it is a potential avenue for future work.

Despite the limitations noted above, the proposed technique performed well on *in vivo* data. Figure 9 shows *in vivo* $R_{2w}^*$ as calculated by three fitting techniques (single-decay, exclusion, dual-decay). The single-decay map clearly overestimates $R_{2w}^*$, and thus cell density, in the tumor. While the exclusion technique exhibits the fewest erroneously fit fat voxels, it simply fits far fewer voxels overall. The *in vitro* data demonstrated that this exclusion technique is not sensitive to voxels with mixed content. Therefore, this map certainly excludes some voxels containing SPIO-labeled cells and would underestimate cell density. In the tumor, the bSSFP image shows a hypo-intense region that is not shown to have high $R_{2w}^*$ by any mapping technique – this is likely necrosis. However, there are hypo-intense regions which have high $R_2^*$ on the single-decay and the dual-decay maps but are nearly ignored by the exclusion technique. The proposed dual-decay technique is less sensitive to misidentification of fat voxels than the single-decay fit while retaining mixed voxels of fat and iron unlike single-decay with post-hoc exclusion.

**Conclusion**

The principal objective of this work was to investigate the adverse effect of contaminant fat modulations in TurboSPI and propose corrective methods. $R_2^*$ estimates should be invariant to fat content, but single-species mono-exponential $R_2^*$ mapping fails in the presence of fat. This is detrimental to *in vivo* cell tracking studies that measure cell density via $R_2^*$ mapping, since mixed-content voxels arise in tumors and lymph nodes surrounded by adipose tissue. *In silico* data shows that $R_2^*$ estimates are stable across most fat fractions if $R_{2w}^*$, $R_{2f}^*$ and fat fraction are estimated simultaneously using a modified dual-decay Dixon model. This finding was corroborated by an *in vitro* experiment in which the proposed model outperformed previous methods when SPIO-labeled cells are present with fat ≥15%. Preliminary *in vivo* results indicate positive development with an improved balance of specificity and sensitivity. Therefore, the proposed model is a promising tool for quantitative TurboSPI $R_2^*$ cell tracking, with further refinements offering the possibility of even higher specificity and sensitivity.

**Acknowledgements**

The authors would like to thank and acknowledge assistance from Christa Davis for imaging mice and Marie-Laurence Tremblay for cell preparation (both in vitro and in vivo).

**Figures**

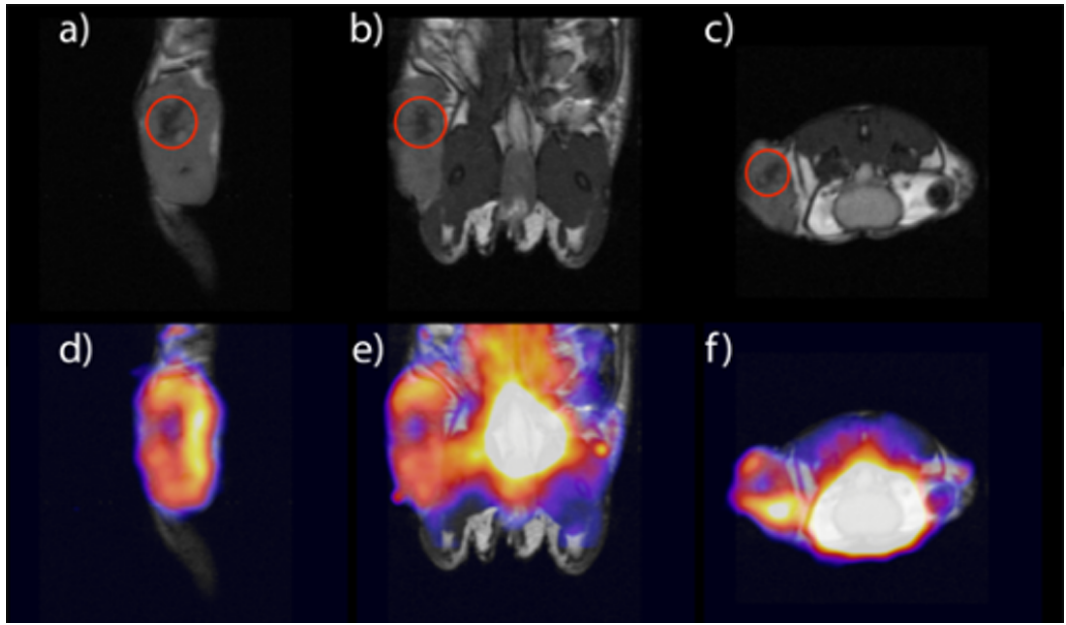

**Figure 1:** Figure reproduced from [12]. Upper row: MRI obtained using bSSFP. Negative contrast in the center of the flank tumor suggests SPIO uptake or necrosis. Bottom row: $^{18}$F-fluorodeoxyglucose (FDG) PET/MRI overlay using simultaneously obtained PET data. A necrotic core is confirmed by low $^{18}$F-FDG activity aligned with the hypo-intense region on the MRI.

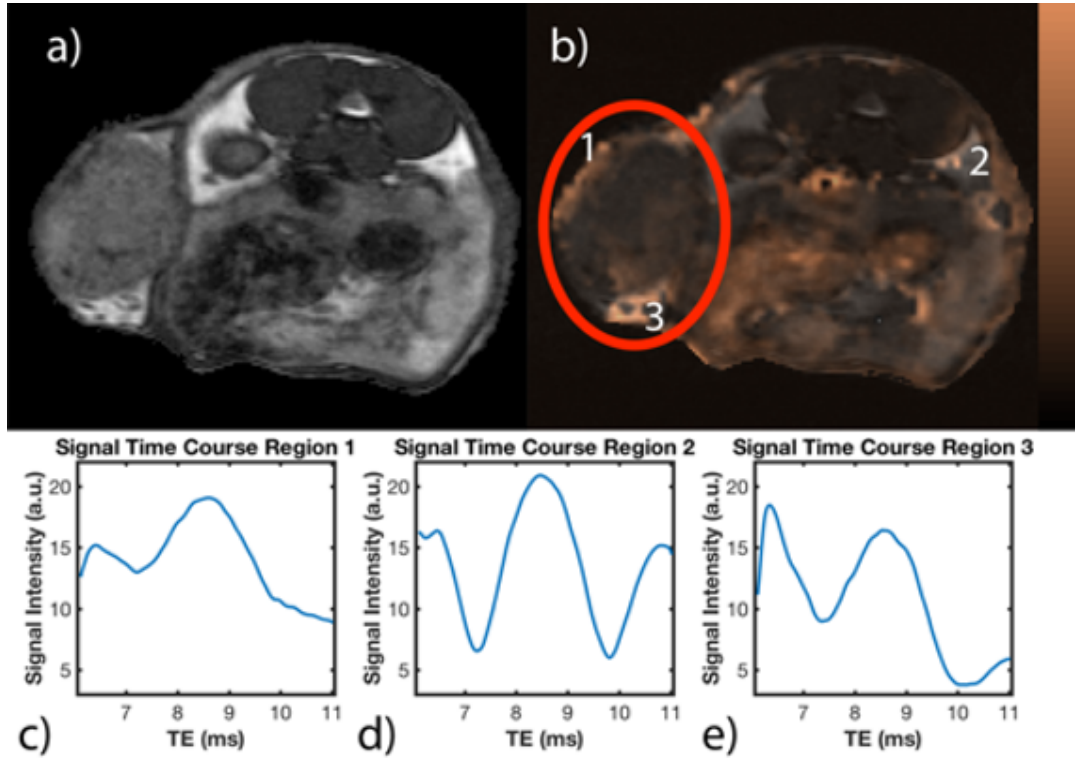

**Figure 2:** Upper row: a) bSSFP MRI. b) $R_2^*$ map (voxel intensity in $0 - 600 s^{-1}$) overlaid on the axial bSSFP slice to show distribution of $R_2^*$. The flank tumour ROI is highlighted by a red circle. Bottom row: c) TurboSPI signal time course from a typical voxel in Region 1, which is a section of tumor potentially containing SPIO and fat. d) A typical time course from Region 2 indicates signal from pure fat (oscillations are fat specific). e) Time courses from Region 3 in the lymph node/adjacent fat pad potentially containing fat and SPIO.

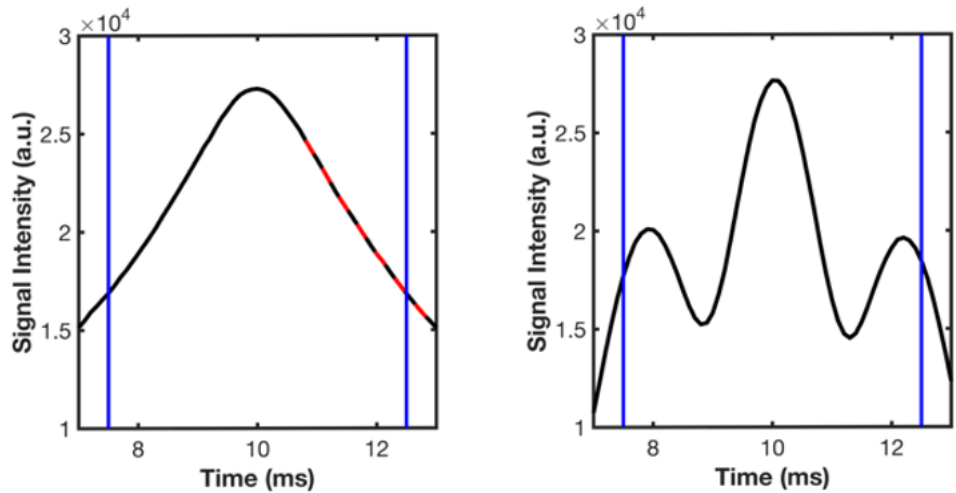

**Figure 3:** *a)* Simulated TurboSPI signal in the presence of SPIO. Dashed red line indicates typical $R_2^*$ fitting region, which avoids non-exponential behavior from

Submitted to NMR in Biomedicine

diffusion near the spin echo peak. *b)* Simulated TurboSPI signal with 15% added fat signal. Off resonance fat causes modulations in the signal time course such that mono-exponential models are not suitable for curve-fitting. Blue lines indicate the experimental acquisition window.

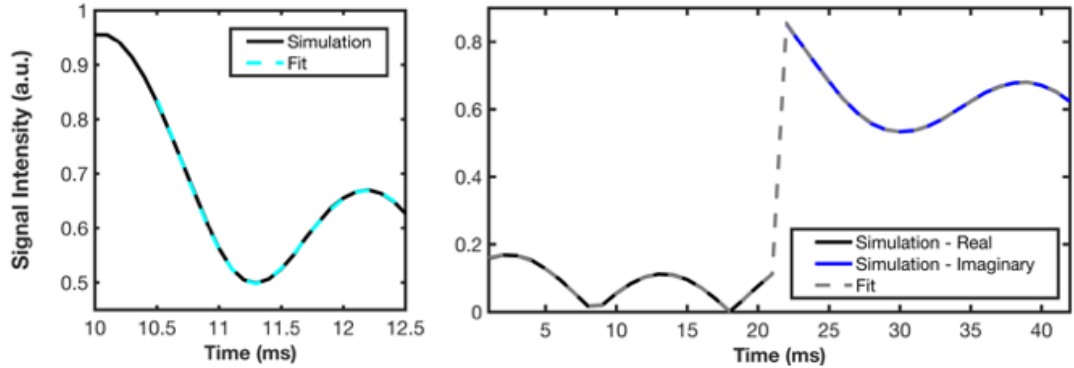

**Figure 4:** a) Normalized simulated complex data (magnitude shown, solid black) with the corresponding fit (dashed blue). The simulation is shown for t = 10 ms to 12.5 ms but the fit is performed for t = 10.5 ms to 12.5 ms to avoid the spin-echo peak. Simulation parameters: Fat fraction ff = 0.15, ζ = 1.5 x 10-5, R = 8um, Δχ = 0.05, no $T_2^*$ effect simulated for the fat species. $T_2^*$ estimate from fit = 4.38 ms, gold standard for same parameters but ff = 0 was $T_2^*$ = 4.5 ms, b) The same simulated data with concatenated real (black) and imaginary (blue) components, and corresponding model fit (dashed grey). Same parameters as in (a) but with a simulated $T_2^*$ = 15 ms for the fat species. Water $T_2^*$ estimate from fit = 4.31 ms.

<="">Submitted to NMR in Biomedicine</>

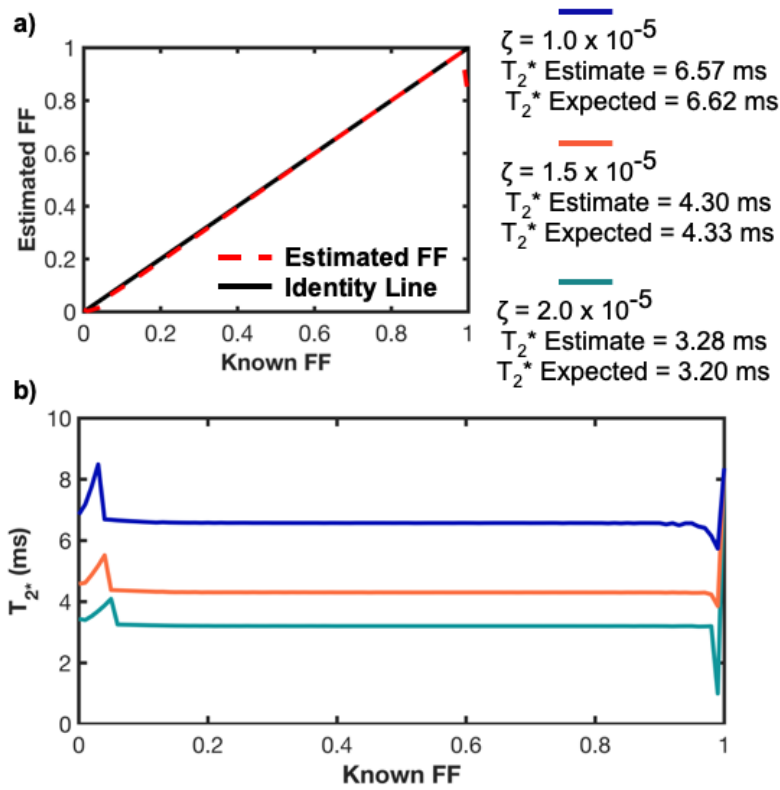

**Figure 5:** a) Fitted fat fraction versus input fat fraction compared to an identity line for reference. Estimates fail only at FF extrema. b) Fitted $T_2^*$ versus input fat fraction shows stability in the measurement except at FF extrema. Decay times change appropriately with volume fraction. Estimates from FF = 0.5

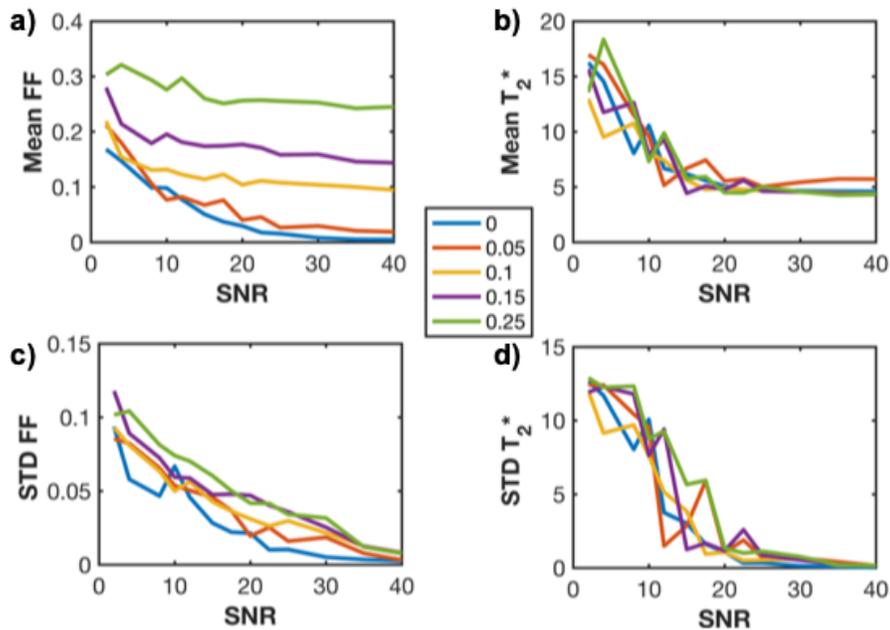

**Figure 6:** Results of fat fraction and $T_2^*$ estimation on simulated data with added Gaussian white noise. Left panels: mean (a) and standard deviation (c) of fat fraction estimates over 20 repetitions, as a function of temporal SNR and for 5 different fat fractions. Right panels: mean (b) and standard deviation (d) of water $T_2^*$ estimates, with gold standard (noiseless, no fat) $T_2^*$ = 4.50ms. In both cases the parameter estimates stabilize above a temporal SNR of 20.

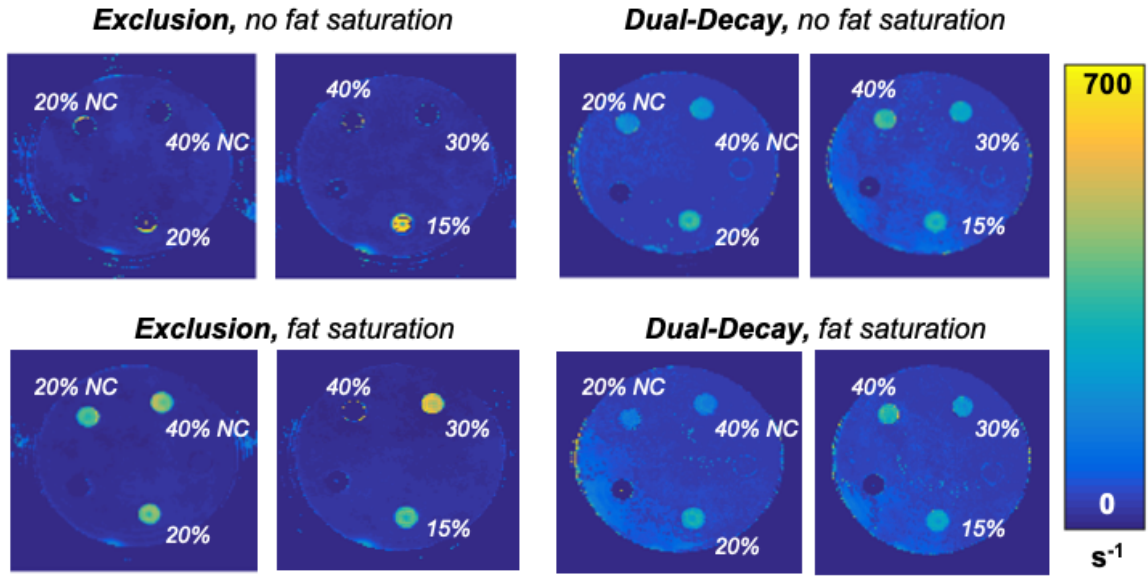

**Figure 7:** Example $R_{2w}^*$ maps for data acquired with and without fat saturation, fitted with the Exclusion and the Dual-Decay methods. All phantoms have equal cell concentration (except for "no cell" (NC) controls) and varying fat fraction (FF). Ideally, NC tubes should have low $R_{2w}^*$, others should have equal $R_{2w}^*$, and fat saturation should not affect $R_{2w}^*$. Estimates from the Dual-Decay technique are stable over a greater range of FF than the Exclusion technique.

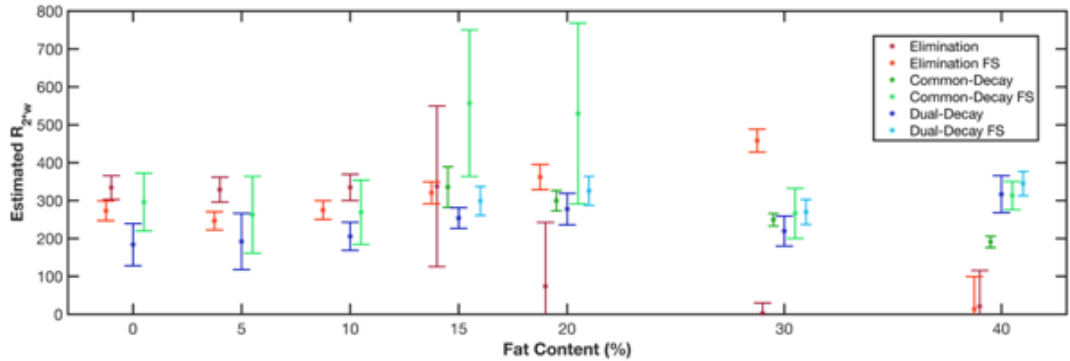

**Figure 8:** $R_{2w}^*$ values for phantoms with equal cell density but varying fat content. Values were estimated using 1) the exclusion method, 2) a common decay rate model ($R_{2w}^* = R_{2f}^*$), and 3) the proposed dual-decay model. Data were acquired with and without a fat saturation pulse. The dual-decay estimates are stable over the widest range of fat content and are least affected by fat saturation. The common decay method results in $R_2^*$ estimates that decrease with fat content, as predicted.

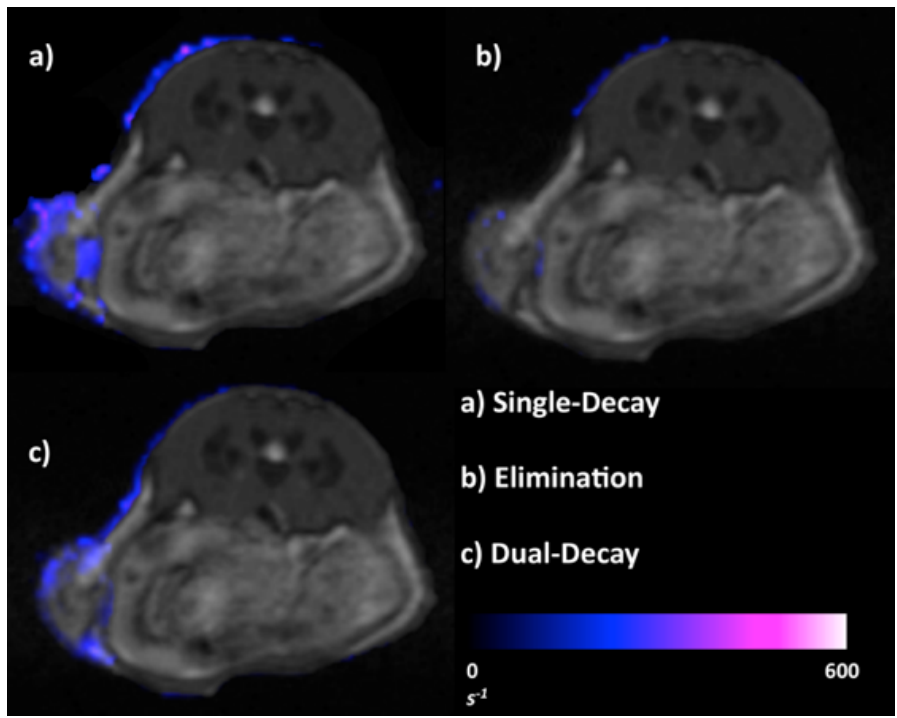

**Figure 9:** Preliminary in vivo demonstration of $R_2^*$ maps (overlays, masked to tumor region) produced by single-decay, exclusion, and the proposed dual-decay fitting techniques. The proposed technique results in fewer erroneously fitted fat voxels than the single-decay technique, but more than exclusion. However, as illustrated *in vitro*, the exclusion technique underestimates the contribution of mixed voxels.

**References**

1. Oiseth SJ, Aziz MS. Cancer immunotherapy: a brief review of the history, possibilities, and challenges ahead. *Journal of Cancer Metastasis and Treatment*. 2017;3(10):250. doi:10.20517/2394-4722.2017.41.

2. Chen DS, Mellman I. Elements of cancer immunity and the cancer–immune set point. *Nature*. 2017;541(7637):321-330. doi:10.1038/nature21349.

3. Eisenhauer EA, Therasse P, Bogaerts J, et al. New response evaluation criteria in solid tumours: revised RECIST guideline (version 1.1). *Eur J Cancer*. 2009;45(2):228-247. doi:10.1016/j.ejca.2008.10.026.

4. Ribas A, Chmielowski B, Glaspy JA. Do We Need a Different Set of Response Assessment Criteria for Tumor Immunotherapy? *Clinical Cancer Research*. 2009;15(23):7116-7118. doi:10.1158/1078-0432.CCR-09-2376.

5. Foster-Gareau P, Heyn C, Alejski A, Rutt BK. Imaging single mammalian cells with a 1.5 T clinical MRI scanner. *Magn Reson Med*. 2003;49(5):968-971. doi:10.1002/mrm.10417.

6. Heyn C, Bowen CV, Rutt BK, Foster PJ. Detection threshold of single SPIO-labeled cells with FIESTA. *Magn Reson Med*. 2005;53(2):312-320. doi:10.1002/mrm.20356.

7. Brewer KD, Lake K, Pelot N, et al. Clearance of depot vaccine SPIO-labeled antigen and substrate visualized using MRI. *Vaccine*. 2014;32(51):6956-6962. doi:10.1016/j.vaccine.2014.10.058.

8. Bernas LM, Foster PJ, Rutt BK. Imaging iron-loaded mouse glioma tumors with bSSFP at 3 T. *Magn Reson Med*. 2010;64(1):23-31. doi:10.1002/mrm.22210.

9. Ribot EJ, Duriez TJ, Trotier AJ, Thiaudiere E, Franconi J-M, Miraux S. Self-gated bSSFP sequences to detect iron-labeled cancer cells and/or metastases in vivo in mouse liver at 7 Tesla. *Journal of Magnetic Resonance Imaging*. 2014;41(5):1413-1421. doi:10.1002/jmri.24688.

10. Majumdar S, Zoghbi SS, Gore JC. The influence of pulse sequence on the relaxation effects of superparamagnetic iron oxide contrast agents. *Magn Reson Med*. 1989;10(3):289-301. doi:10.1002/mrm.1910100302.

11. Wang PC, Shan L. Essential Elements to Consider for MRI Cell Tracking Studies with Iron Oxide-based Labeling Agents. *J Basic Clin Med*. 2012;1(1):1-6.

12. O'Brien-Moran Z, Tremblay M-L, Davis C, Rioux JA, Brewer KD. Improved Tracking and Quantification of SPIO-Labeled Cells Using bSSFP with Compressed Sensing TurboSPI. In: Honolulu; 2017.

13. Brewer KD, Stanley O, Davis C, et al. Tracking SPIO-Labeled Effector & Regulatory Cell Migration with MRI. In: Savannah; 2013.

14. Wang Q, Zhang G, Li K, Quan Q. R2* and R2 mapping for quantifying recruitment of superparamagnetic iron oxide-tagged endothelial progenitor cells to injured liver: tracking in vitro and in vivo. *International Journal of Nanomedicine*. 2014;9(1):1815-1822. doi:10.2147/IJN.S58269.

15. van Buul GM, Kotek G, Wielopolski PA, et al. Clinically translatable cell tracking and quantification by MRI in cartilage repair using superparamagnetic iron oxides. Tjwa M, ed. *PLoS ONE*. 2011;6(2):e17001. doi:10.1371/journal.pone.0017001.

16. Rioux JA, Beyea SD, Bowen CV. 3D single point imaging with compressed sensing provides high temporal resolution R 2* mapping for in vivo preclinical applications. *MAGMA*. August 2016. doi:10.1007/s10334-016-0583-y.

17. Rioux JA, Brewer KD, Beyea SD, Bowen CV. Quantification of superparamagnetic iron oxide with large dynamic range using TurboSPI. *J Magn Reson*. 2012;216:152-160. doi:10.1016/j.jmr.2012.01.017.

18. Srinivas M, Heerschap A, Ahrens ET, Figdor CG, de Vries IJM. (19)F MRI for quantitative in vivo cell tracking. *Trends Biotechnol*. 2010;28(7):363-370. doi:10.1016/j.tibtech.2010.04.002.

19. Srinivas M, Boehm-Sturm P, Figdor CG, de Vries IJ, Hoehn M. Labeling cells for in vivo tracking using 19F MRI. *Biomaterials*. 2012;33(34):8830-8840. doi:10.1016/j.biomaterials.2012.08.048.

20. Fox MS, Gaudet JM, Foster PJ. Fluorine-19 MRI Contrast Agents for Cell Tracking and Lung Imaging. *Magnetic Resonance Insights*. 2015;8(Suppl 1):53-67. doi:10.4137/MRI.S23559.

21. Beyea SD, Balcom BJ, Prado PJ, et al. Relaxation Time Mapping of ShortT*2Nuclei with Single-Point Imaging (SPI) Methods. *Journal of Magnetic Resonance*. 1998;135(1):156-164. doi:10.1006/jmre.1998.1537.

22. Beyea SD, Balcom BJ, Mastikhin IV, Bremner TW, Armstrong RL, Grattan-Bellew PE. Imaging of Heterogeneous Materials with a Turbo Spin Echo Single-Point Imaging Technique. *Journal of Magnetic Resonance*. 2000;144(2):255-265. doi:10.1006/jmre.2000.2054.

23. Yu H, Shimakawa A, McKenzie CA, Brodsky E, Brittain JH, Reeder SB. Multiecho water-fat separation and simultaneous R2* estimation with multifrequency fat spectrum modeling. *Magn Reson Med*. 2008;60(5):1122-1134. doi:10.1002/mrm.21737.

24. O'Regan DP, Callaghan MF, Wylezinska-Arridge M, et al. Liver Fat Content and T2*: Simultaneous Measurement by Using Breath-hold Multiecho MR Imaging at 3.0 T—Feasibility. *Radiology*. 2008;247(2):550-557. doi:10.1148/radiol.2472070880.

25. Chebrolu VV, Hines CDG, Yu H, et al. Independent estimation of T*2 for water and fat for improved accuracy of fat quantification. *Magn Reson Med*. 2010;63(4):849-857. doi:10.1002/mrm.22300.

26. Reeder SB, Cruite I, Hamilton G, Sirlin CB. Quantitative assessment of liver fat with magnetic resonance imaging and spectroscopy. *Journal of Magnetic Resonance Imaging*. 2011;34(4):729-749. doi:10.1002/jmri.22580.

27. Glover GH. Multipoint dixon technique for water and fat proton and susceptibility imaging. *Journal of Magnetic Resonance Imaging*. 1991;1(5):521-530. doi:10.1002/jmri.1880010504.

28. Horng DE, Hernando D, Hines CDG, Reeder SB. Comparison of R2* correction methods for accurate fat quantification in fatty liver. *Journal of Magnetic Resonance Imaging*. 2013;37(2):414-422. doi:10.1002/jmri.23835.

29. Bowen CV, Zhang X, Saab G, Gareau PJ, Rutt BK. Application of the static dephasing regime theory to superparamagnetic iron-oxide loaded cells. 2002;48(1):52-61. doi:10.1002/mrm.10192.

30. Kiselev VG, Posse S. Analytical model of susceptibility-induced MR signal dephasing: Effect of diffusion in a microvascular network. *Magn Reson Med*. 1999;41(3):499-509. doi:10.1002/(SICI)1522-2594(199903)41:3<499::AID-MRM12>3.0.CO;2-O.

31. Feltkamp MC, Smits HL, Vierboom MP, et al. Vaccination with cytotoxic T lymphocyte epitope-containing peptide protects against a tumor induced by human papillomavirus type 16-transformed cells. *Eur J Immunol*. 1993;23(9):2242-2249. doi:10.1002/eji.1830230929.

32. Smith KA, Meisenburg BL, Tam VL, et al. Lymph node-targeted immunotherapy mediates potent immunity resulting in regression of isolated or metastatic human papillomavirus-transformed tumors. *Clinical Cancer Research*. 2009;15(19):6167-6176. doi:10.1158/1078-0432.CCR-09-0645.

# Cell Density Quantification with TurboSPI: $R_2^*$ Mapping with Compensation for Off-Resonance Fat Modulation

Zoe O'Brien-Moran[1,2], Chris V. Bowen[1,2], James A. Rioux[1,2], Kimberly D. Brewer[1,2]

**Supplementary Materials**

Methods

For previous experiments cited in [22] and shown in Figures 1 and 2 as supporting data, all *in vivo* data was acquired using the same C3 cancer model and injected CTLs as described in methods. Mice were imaged on a 3.0T preclinical MRI system with the same bSSFP sequence used as an anatomical reference as described for the other in vivo experiments. The PET data shown in Figure 1 was acquired at the same time as the MRI data using a nuPET insert (2 rings with silicon photomultiplier detectors; Cubresa, Winnipeg, MB) for the MRI system. The RF coil used for this experiment was a mouse body quad coil. Approximately 600uCi of [18]F-fluorodeoxyglucose (FDG) was injected in the tail vein of the mouse prior to imaging. PET imaging was done simultaneously to MRI, with PET data acquisition beginning 50 minutes post-injection. The PET scan lasted for 30 minutes. PET data was reconstructed with a iterative maximum likelihood estimate OSMAPOSL algorithm (Cubresa).

Results

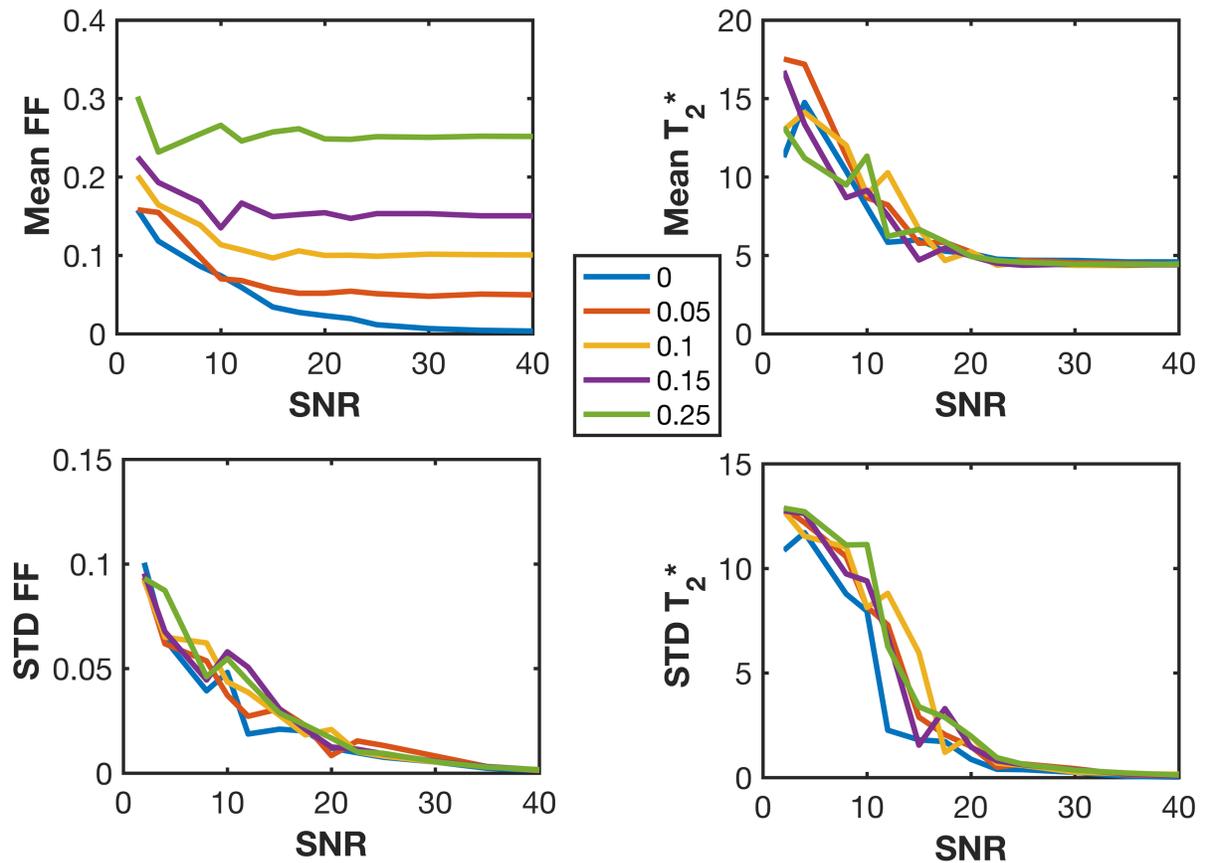

***Supplementary Figure 1*** – Results of fat fraction and $T_2^*$ estimation on simulated data with added Gaussian white noise for $T_{2F}^* = T_{2w}^*$. Left panels: mean (a) and standard deviation (c) of fat fraction estimates over 20 repetitions, as a function of temporal SNR and for 5 different fat fractions. Right panels: mean (b) and standard deviation (d) of water $T_2^*$ estimates, with gold standard (noiseless, no fat) $T_2^*$ = 4.50ms. In both cases the parameter estimates stabilize above a temporal SNR of 20.